\documentclass[conference]{IEEEtran}
\usepackage{cite}
\usepackage{amsmath,amssymb,amsfonts}
\usepackage{algorithmic}
\usepackage{graphicx}
\usepackage{textcomp}
\usepackage{xcolor}
\usepackage[capitalise]{cleveref}
\usepackage{balance}

\def\BibTeX{{\rm B\kern-.05em{\sc i\kern-.025em b}\kern-.08em
    T\kern-.1667em\lower.7ex\hbox{E}\kern-.125emX}}
\begin{document}

\title{Unit Testing Past vs. Present: Examining \\
LLMs' Impact on Defect Detection and Efficiency}

\author{\IEEEauthorblockN{Rudolf Ramler}
\IEEEauthorblockA{
\textit{Software Competence} \\
\textit{Hagenberg GmbH (SCCH)} \\
Hagenberg, Austria \\
}
\and
\IEEEauthorblockN{Philipp Straubinger}
\IEEEauthorblockA{
\textit{University of Passau}\\
Passau, Germany \\
}
\and
\IEEEauthorblockN{Reinhold Plösch}
\IEEEauthorblockA{
\textit{Johannes Kepler University Linz}\\
Linz, Austria \\
}
\and
\IEEEauthorblockN{Dietmar Winkler}
\IEEEauthorblockA{
\textit{Austrian Center for Digital}\\
\textit{Production and TU Wien}\\
Vienna, Austria \\
}
}

\maketitle

\begin{abstract}
The integration of Large Language Models (LLMs), such as ChatGPT and GitHub Copilot, into software engineering workflows has shown potential to enhance productivity, particularly in software testing. This paper investigates whether LLM support improves defect detection effectiveness during unit testing. Building on prior studies comparing manual and tool-supported testing, we replicated and extended an experiment where participants wrote unit tests for a Java-based system with seeded defects within a time-boxed session, supported by LLMs. Comparing LLM supported and manual testing, results show that LLM support significantly increases the number of unit tests generated, defect detection rates, and overall testing efficiency. These findings highlight the potential of LLMs to improve testing and defect detection outcomes, providing empirical insights into their practical application in software testing.
\end{abstract}


\section{Introduction}

Large Language Models (LLMs) in form of ChatGPT, Copilot, Gemini, etc. are now commonly used for everyday tasks. They are also increasingly integrated into software engineering workflows, offering new ways to enhance productivity and accuracy~\cite{DBLP:journals/ese/ZhengNZCCGWW25}. While their applications span diverse areas like requirements analysis, code generation, documentation, and debugging, their role in software testing—particularly unit testing—has garnered widespread interest. Recent studies suggest that LLMs can assist in generating test cases, improving coverage, and enhancing defect detection~\cite{DBLP:conf/kbse/YangYGW0ZCZLW024}.
Most of these studies, however, focus on applying LLMs for automating the test case generation process, while in practice testers are interactively using LLMs to support all kind of tasks during unit testing. Thus, we aim to investigate the broad question: \textit{Do LLMs support human testers in finding defects in unit testing?}

In previous research, we compared tool-supported and manual unit testing approaches in controlled experiments~\cite{DBLP:conf/euromicro/RamlerWS12}. 
Manual unit testing was evaluated through a time-constrained competition among students, who aimed to identify as many defects as possible in a Java-based system with seeded defects. The findings revealed that the participants were able to detect an average of 3.71 defects in 60 minutes testing, showing varying levels of efficiency. Subsequent replication of the experiment with professional developers demonstrated no significant differences in the results, validating the use of student participants as a proxy for practitioners~\cite{DBLP:conf/compsac/RamlerWK13}.

Building on this foundation, we investigate the impact of LLMs on defect detection during unit testing by repeating and extending the original study. Our experiment evaluates whether integrating LLMs enhances participants' ability to detect defects under similar time and resource constraints. By analyzing the created test cases and their defect detection performance, we aim to provide empirical evidence on the efficacy of LLM supported unit testing, where LLMs are used interactively in the loop of human testing activities. 

Our study with 30 participants shows that LLM-supported unit testing significantly outperforms manually writing unit tests in both productivity and defect detection. Participants using LLMs created more tests, achieved higher coverage, and identified more defects. However, the increase in test quantity also led to a higher rate of false positives, highlighting a trade-off between productivity and precision.
Given that the original study was conducted over a decade ago, our comparison offers an interesting insight into how unit testing practices have evolved. After many years of relatively small and stable advancements, the advent of LLMs appears to have had the most significant impact on unit testing.

\section{Experiment Design}

With the research objective outlined in the introduction, the experiment was conducted following the experimental design established in the original studies~\cite{DBLP:conf/euromicro/RamlerWS12} and \cite{DBLP:conf/compsac/RamlerWK13}.

\subsection{Study Material}

The materials for the experiment include the system under test (SUT) with seeded defects from the original study~\cite{DBLP:conf/euromicro/RamlerWS12}. Additionally, we created detailed instructions for the participants and a post-experiment questionnaire to complete after the testing task.

The SUT consists of a Java collection class library similar to the standard Java Collections Framework. It comprises about 2,800 lines of code in 34 classes and interfaces with 164 methods, includes algorithmically complex implementations, and a variety of object-oriented features. Its familiarity to participants and prior successful uses in experimentation make it well-suited for our study.
The SUT was provided as a folder containing a JAR archive with the compiled binary files of the collection classes to be tested, the corresponding Javadoc documentation for these classes, and a Maven test project for participants to create and execute their unit tests. The source code of the collection classes was not part of the SUT.

The post-experiment questionnaire collected information on participants' prior experience with unit testing and LLM-based tools. It also inquired whether LLM support was utilized during the experiment (optional) and, if so, which tools were used and how participants assessed their usefulness.

\subsection{Participants}

The participants for the experiment were recruited from a master's-level software testing course. As part of the course, students had gained both theoretical knowledge and practical experience in testing, including assignments that focused on writing unit tests with the JUnit framework. Additionally, many students had prior knowledge of unit testing and practical experience in software development. 

Participation in the experiment was voluntary, with all results anonymized for evaluation. Participants were given the option to exclude their data from the study. Of the 34 students who participated, 3 chose to opt out, leaving data from 31 participants available for analysis. 

\subsection{Execution}

The experiment was offered as an optional exercise within the software testing course, with all participants receiving extra points as an incentive. Each participant was required to complete the task individually within a 60-minute time limit.

Participants received the following instructions: \textit{"Write JUnit tests for the collection classes that represent the system under test, which contains defects. You may use LLM support or LLM-based tools of your choice for this task; however, their use is optional. Your goal is to identify as many defects as possible. For each defect found, provide a failing unit test that exposes the defect. At the end of the experiment, submit all tests you wrote—both failing and passing. The time allotted for writing tests is strictly limited to 60 minutes."}

Upon completing the task, participants submitted a folder containing the source code of the unit tests they created, along with the completed questionnaire.

\subsection{Analysis and Evaluation}

All submissions were anonymized, verified for completeness and consistency, and prepared for automated processing. Data from participants who opted out was excluded from the evaluation. Tests that failed to compile were commented out and excluded from further analysis. 

To identify false positives—tests failing despite no seeded defects being present—all tests were executed against the clean, defect-free version of the SUT. False positive tests were excluded from subsequent runs using the JUnit annotation \textit{@Disable}. Prior to exclusion, each failing test was manually inspected by a researcher to confirm its classification as a false positive. For instance, tests were marked as false positives when their assertions contradicted the documented behavior in the associated Javadoc.
A total of 158 failing tests were inspected, during which three instances revealed a new, previously unknown defect in the SUT. This newly discovered defect was included in the defect counts for subsequent analysis. 

To evaluate defect detection, the test suites of all 31 participants were executed against 37 defective versions of the SUT, each containing a specific seeded defect. A defect was marked as detected if one or more test cases failed during a run. In total, 1,147 test runs were conducted (31 test suites executed against 37 defective versions of the SUT).

\section{Results and Discussion}
\begin{figure*}[t]
\includegraphics[width=\textwidth]{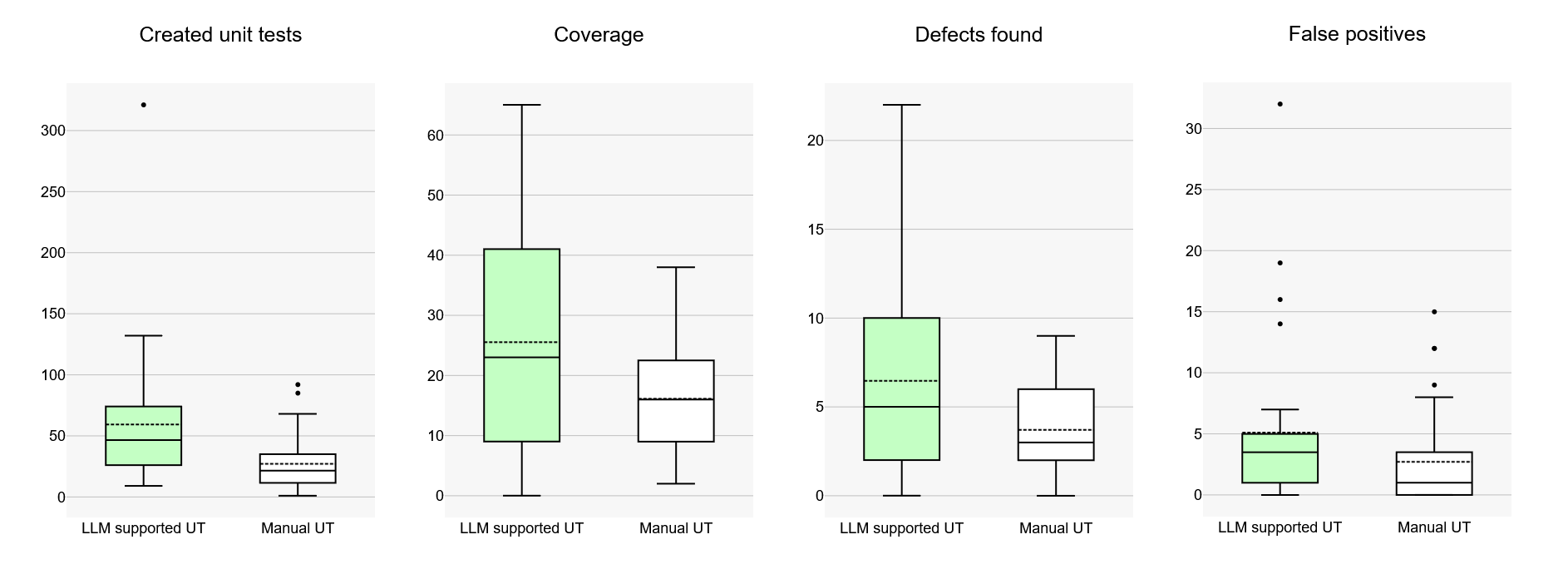}
\caption{Comparison of Experiment Group and Control Group.}
\label{fig_comparison_combined}
\end{figure*}

This section presents the results of our experiment on unit testing with LLM support, and it provides a comparative analysis with the findings of the previous study on participants performing manual unit testing \cite{DBLP:conf/euromicro/RamlerWS12}, which serves as control group. 
The comparison is based on the \textit{number of created unit tests}, the \textit{coverage achieved} by the tests, the \textit{number of defects found}, and the \textit{number of false positives}. The plots in Fig.~\ref{fig_comparison_combined} show these numbers side by side. An overview of all metrics computed for our analysis is provided in~\cref{table1}. 

\begin{table}[t]
\renewcommand{\arraystretch}{1.3}
\caption{Results from LLM Supported Unit Testing (this experiment) and Comparison to Manual Unit Testing (control group)}
\label{table1}
\begin{center}
\begin{tabular}{|l|p{2cm}|p{2cm}|}
\hline
& \textbf{LLM supported} & \textbf{Manual} \\
& \textbf{unit testing} & \textbf{unit testing~\cite{DBLP:conf/euromicro/RamlerWS12}} \\ 
\hline
\multicolumn{3}{|l|}{\textbf{Study participants}} \\
\hline
total (after opt-out) & 30 & 48 \\
\hline
\multicolumn{3}{|l|}{\textbf{Created unit tests}} \\
\hline
total (all participants) & \textbf{1780} & 1301 \\
average / median & \textbf{59.3 / 46.5} & 27.1 / 21.5\\
min--max & \textbf{9--321} & 1--92 \\
\hline
\multicolumn{3}{|l|}{\textbf{Coverage (branches)}} \\
\hline
total (all participants) & \textbf{74\%} & 67\% \\
average / median & \textbf{26\% / 23\%} & 16\% / 16\% \\
min--max & \textbf{0--65\%} & 2--38\% \\
\hline\multicolumn{3}{|l|}{\textbf{Defects found}} \\
\hline
total distinct (all tests) & \textbf{33} & 24 \\
average / median & \textbf{6.5 / 5} & 3.7 / 3\\
min--max & \textbf{0--22} & 0--9 \\
times zero defects found & \textbf{1} & 7 \\
top 10 participants & \textbf{31} & 24 \\
\hline
\multicolumn{3}{|l|}{\textbf{False positives}} \\
\hline
total (all participants) &\textbf{153} & 130 \\
average / median & \textbf{5.1 / 3.5} & 2.7 / 1\\
min--max & \textbf{0--32} & 0--15 \\
times zero false positves & \textbf{3} & 16 \\
top 10 participants &\textbf{78} & 23 \\
\hline
\end{tabular}
\end{center}
\end{table}

\subsection{LLM Support Used for Unit Testing}
After completing the experiment, participants were asked in a questionnaire to report the LLM tools they used for creating unit tests. Participants could select multiple options. The most frequently used tool was ChatGPT, with 15 mentions (33\%) for versions 4/4o and 13 mentions (31\%) for version 3.5. GitHub Copilot was cited 11 times (26\%), while three participants  (7\%) reported using other tools, such as Codium AI or self-hosted models. One participant indicated that no LLM support (\textit{none}) was used. Data from this participant was excluded from the analysis, as the study focused specifically on unit testing with LLM support.

\subsection{Number of Created Unit Tests}

The test suites submitted by the 30 participants (excluding one who did not use LLM support) contained a total of 1,780 executable unit tests. On average, participants utilizing LLM support created 59.3 unit tests within the 60-minute time frame (median = 46.5). The number of tests created ranged from a minimum of 9 to a maximum of 321 unit tests per participant.

In contrast, the control group from the previous study~\cite{DBLP:conf/euromicro/RamlerWS12}, which relied on manual unit testing only, produced significantly fewer tests in the same time (Mann Whitney U-Test, p-value~= .00021). The average was 27.1 tests per participant (median = 21.5), with the best result being 92 tests.

The results demonstrate a \textbf{substantial productivity increase} in the group \textbf{using LLM support}, with the average number of tests per participant more than doubling (+119\%) compared to manual testing.

\subsection{Coverage}

We assessed the branch coverage achieved by the participants' unit tests using JaCoCo. The combined execution of all tests from all participants achieved an overall branch coverage of 74\% across the 292 possible branches of the SUT. The average branch coverage per participant was 26\% (median = 23\%). One participant, who created 31 tests, achieved no coverage (min = 0\%) because the tests inadvertently targeted similar but unrelated collection classes from the Java Development Kit instead of the SUT. The maximum branch coverage achieved by a single participant was 65\%, which was also the participant who created the highest number of tests.

The results indicate that a larger number of tests correlates with higher coverage. To confirm this, we calculated the Spearman correlation 
between the number of tests created per participant and the corresponding coverage. The analysis revealed a strong positive correlation (r = 0.78, p \textless 0.001).

When comparing these findings with the control group from the previous study, we observed that in the current LLM-supported experiment the participants created more unit tests (as described above), which resulted in a significantly higher branch coverage. In the control group, the combined tests of all participants achieved 67\% branch coverage. On average, participants who manually created their tests reached only 16\% branch coverage (median = 16\%), with individual results ranging from 2\% (min) to 38\% (max).

These findings suggest that \textbf{LLM support enabled participants to generate more unit tests}, and the \textbf{increased number of tests} directly contributed to achieving \textbf{higher branch coverage}.

\subsection{Defects Found}

The system under test (SUT) used in the experiment contained 38 known defects: 37 seeded defects from prior studies plus one previously unknown defect identified during this study. Collectively, the 30 participants in the LLM-supported group discovered 33 of these defects. On average, participants in this group identified 6.5 defects (median = 5), with results ranging from a minimum of 0 defects (one participant) to a maximum of 22 defects found by a single participant.

The \textbf{defect detection rate was significantly higher for the LLM-supported group} compared to the control group (Mann Whitney U-Test, p-value~= .01287). In the control group, which relied on manual unit testing, a total of 24 defects were identified, averaging 3.7 defects per participant. Additionally, the range of defects found per participant differed notably as 7 participants in the control group failed to detect any defects, while the top three participants each identified 9 defects.

To further analyze the performance, we compared the results of the top 10 participants from each group. The 10 highest-performing participants with LLM support identified 31 of the 38 known defects, whereas the top 10 participants in the control group found only 24 defects.

This comparison based on ranks helps to mitigate a potential bias caused by participants opting out of data sharing when not performing well, as participation in the experiment was optional, and students could withhold their data even if they completed the experiment. A possible indicator of such bias is the fact that only one participant in the LLM-supported group found no defects, compared to seven participants in the control group who failed to detect any defects. 

Even when results are analyzed by rank 
, participants using LLM support demonstrate superior performance. The \textbf{top-ranked participants with LLM support identified more defects}, and there were fewer lower-ranked participants in this group who found few or no defects. 



\subsection{False Positive Tests}


The notable effort required to handle false positives is of practical relevance and should therefore also be considered in the evaluation. 
On average, participants in the experiment group created 5.1 false positive tests (median = 3.5), with the number of false positives per participant ranging from 0 (3 participants) to a maximum of 32. In comparison, the \textbf{control group produced fewer false positives}, averaging 2.7 per participant (median = 1), with a maximum of 15 false positives. Notably, a substantial proportion of the control group (16 participants) did not generate any false positives.

We further analyzed the false positives generated by the top 10 participants in each group, which we previously identified based on the number of defects they detected. The top 10 participants in the experiment group collectively produced 78 false positives, whereas the control group produced only 23.

To explore the difference in false positives further, we performed a Spearman correlation 
analysis on the experiment group data. The results revealed a strong positive correlation between the number of tests created and the number of false positives (r~= 0.5, p~= .002), indicating that participants who created more tests also tended to produce more false positives.

Since the average number of tests created by participants in the experiment group was more than twice that of the control group (see above), we conclude that the \textbf{increase in average false positives} per participant from 2.7 to 5.1 in the experiment group is primarily \textbf{attributable to the higher number of tests} created. Importantly, we found no evidence to suggest that the use of LLM support influenced the number of false positives.
  
%

\section{Threats to Validity}  
This section summarizes key threats to validity and the measures taken to mitigate them.

\textit{Study participants and experience.} The study involved 31 master's students from a software testing course, most of whom were part-time professionals in software engineering or related fields. A similar study with industry practitioners showed comparable results~\cite{DBLP:conf/compsac/RamlerWK13}.
\textit{Study Process and Duration.} We followed the methodology from the prior study~\cite{DBLP:conf/euromicro/RamlerWS12}, with a 60-minute time limit for creating test cases and identifying defects. The replication study~\cite{DBLP:conf/compsac/RamlerWK13} indicates that testing results may improve if this duration is extended.
\textit{Study Objects.} The SUT comprised Java familiar collection classes similar to those in the Java Development Kit. The familiarity of these classes allows participants to focus on testing tasks but might also positively influence LLM performance.
\textit{Data Collection and Analysis.} Our current study included fewer participants due to offering an opt-out option. This option may have introduced a bias, as low performing participants may have decided to omit their results and only the good results remained. Data analysis followed established methods from previous studies~\cite{DBLP:conf/euromicro/RamlerWS12,DBLP:conf/compsac/RamlerWK13}.

\section{Conclusions and Future Work}

In this study, we repeated an experiment from a study conducted over a decade ago, transferring it to a contemporary setting with LLM support in place. The aim was to investigate (a) the impact of LLM technology on unit testing and defect detection, and (b) to investigate the evolution of unit test development over the last years.

The main findings of this study are twofold:
(a) LLMs can effectively support human testing activities, showing a significant increase in the creation of unit tests and in finding defects.
(b) The introduction of LLMs has a noticeable impact on unit testing efficiency, leading probably to the biggest practical improvement over the past decade.


\textbf{Future work.} We plan to expand this replication across different locations with a larger participant pool. Additionally, we aim to focus more specifically on LLM tools and their support for testing tasks performed by humans to better understand the impact and potential of LLM technologies to change unit testing processes in future.

\section*{Acknowledgment}

This work has been partially funded by the Austrian Research Promotion Agency (FFG) via the COMET competence center INTEGRATE by SCCH (FFG grant no. 892418) and the Austrian Competence Center for Digital Production (CDP, FFG grant no. 892418). Finally, we thank all students for participating in the experiment and sharing their data.

\balance
\bibliographystyle{ieeetr}
\bibliography{bib}

\end{document}